\documentclass[fleqn,twoside]{article}
\usepackage{espcrc2}

\usepackage{epsfig}
\usepackage[figuresright]{rotating}

\newcommand{\beq}{\begin{equation}}
\newcommand{\eeq}{\end{equation}}

\newcommand{\AmS}{{\protect\the\textfont2 A\kern-.1667em\lower.5ex\hbox{M}\kern-.125emS}}
\newcommand{\beqa}{\begin{eqnarray}}
\newcommand{\eeqa}{\end{eqnarray}}

\newcommand{\mysection}[1]{\setcounter{equation}{0}\section{#1}}

\def\blft{\hspace*{-0.4cm}}
\title{Dynamical Zero Modes and Criticality in Continuous Light Cone Quantization of $\Phi^{4}_{1+1}$  }

\author{P. Grang\'e\address[MCSD]{ Laboratoire de Physique Math\'ematique et Th\'eorique,
          Universit\'e Montpellier II, \\ 
        F-34095 Montpellier Cedex 05, France}%
        \thanks{CNRS-UMR 5825 },
        S. Salmons\addressmark,
        E. Werner\address{Institut f\"ur Theoretische Physik
          Universit\"at Regensburg, \\ D-93040 Regensburg, Germany
       }}
       
\begin{document}

\begin{abstract}
Critical behaviour of the $2D$ scalar field theory in the
LC framework is reviewed. The notion of  dynamical zero modes is introduced and shown
 to lead to a non trivial covariant dispersion relation only for Continuous LC Quantization (CLCQ). The critical exponent $\eta$ is found to be governed by the behaviour of 
the infinite volume limit under conformal transformations properties preserving
 the local LC structure. The $\beta$-function is calculated exactly and found non-analytic, with a critical exponent $\omega=2$, in agreement with the conformal field theory prediction of Calabrese et al.
\vspace{1pc}
\end{abstract}

\maketitle
\renewcommand{\thepage}{\arabic{page}}
\mysection{Introduction}\label{Introduction}

It is by now well established that the infrared behavior of the \( \Phi ^{4}_{1+1} \) LC field theory
 is controlled by a constrained zero mode of the field operator,
which is the LC signature of a nontrivial vacuum. This constrained zero mode
appears if the dimensionless coupling $g$ is stronger than a critical coupling
$g_{cr}$. For $g > g_{cr} $ the order parameter $\left\langle 0\left| \Phi \right| 0\right\rangle$
becomes nonzero. The characteristics of the resulting second order phase transition
are those of a mean field theory. The constrained zero mode being a static quantity
it is not able to furnish information on the dynamics of the phase transition
as for example: fluctuations of the order parameter, long range correlation functions, 
dispersion relations for the fluctuating fields etc....
All these phenomena are conditioned by ``Dynamical Zero Modes'' which we shall
describe and study in the following. \emph{In doing this we shall make an important simplification:}
Since the dynamics of the phase transition is governed by long range fluctuations
whose amplitudes are small close to the transition point, we can linearize the
equations of motion of the fluctuating fields.

\mysection{Field Decomposition}

The Lagrangien  $L = \frac{1}{2} \left( \partial ^{+}\Phi \right) \left( \partial ^{-}\Phi \right)
- \frac{1}{2} m^{2} \Phi ^{2} - \frac{\lambda }{4!}\Phi ^{4}$ leads to the equation of motion (EQM) 
$\partial ^{+}\partial ^{-}\Phi - m^{2} \Phi - \frac{\lambda }{3!}\Phi ^{3}=0.$\\
We decompose the field  $\Phi \left( x\right)$ into a classical field  $\Phi _{c}\left( x\right)$
and a quantum field $\Psi \left( x\right)$ : $\Phi \left( x\right) = \Phi _{c}\left( x\right) + \Psi \left( x\right).$ The classical field acts as a kind of background for the quantum field: it is
tantamount to an inhomogenous medium in which the quantum field evolves.
This implies a generalization of the Haag series for the field $\Psi \left( x\right)$ :\\

\blft$\Psi \left( x\right)=\varphi _{0}\left( x\right) + \int dx_{1}^{-}dx_{2}^{-}[\,: \varphi _{0} \left( x_{1}\right) \varphi _{0}\left( x_{2}\right) : $
\beq
\label{Haagser} 
g_{2} \left( x^{-}_{1}-x^{-}, x^{-}_{2}-x^{-}; x\right)\,]+ \psi _{3}\left( x\right) + .........
\eeq 
with $x=\left( x^{-},x^{+}\right).$ The second term on the r.h.s. of
Eq. (\ref{Haagser}) contains two free-field operators \( \varphi _{0}\left( x\right)  \)
and will be abbreviated by \( \psi _{2}\left( x\right)  \), the next one $\psi _{3}\left( x\right)$
 contains three free-field operators, etc.
The free-field $\varphi_{0} \left( x\right)$ has the Fock-expansion \cite{GUW}:\\

\blft $\varphi_0(x)=\int_{0}^{\infty} {dk^+ \over {\sqrt{4 \pi k^+}}} f(k^+)$
\beq
\label{Fex}
\left[ a\left( k^{+}\right) e^{\frac{i}{2}k x}+a^{+}\left( k^{+}\right) e^{-\frac{i}{2}k x}\right],
\eeq
with $\left[ a\left( k^{+}_{1}\right), a^{+}\left( k^{+}_{2}\right)\right]=\delta \left( k^{+}_{1}-k^{+}_{2}\right)$
and $k.x=\hat{k}^{-} x^{+}+k^{+} x^{-}$, $\hat{k}^{-}=\frac{m^2}{k^{+}}$ $(m^2 > 0).$ The properties of the test 
function $f(k^{+})$ are discussed in \cite{GUW} : its effect is to render the integral finite at both ends of the
spectrum in $k^{+}$ and the parameter entering any possible form of  $f(k^{+})$ is related in an essential way to
the final renormalization procedure of the theory.

Some remarks are in order:

(1) The contributions $\psi _{2}$ and $\psi _{3}$ are of different
character: while $\psi_{3}$ has perturbative contributions, $\psi _{2}$
is purely nonperturbative and does arise only in the presence of constrained
or dynamical zero modes.

(2) The constrained zero mode is obtained from projection of  $\psi _{2}$ onto the
 vacuum sector.

For the following we need the detailed structure of $\psi_{2}\left( x\right)$ :\\

\blft$\psi_{2}(x) = \int_{0}^{\infty}{{dk^+_1 dk^+_2} \over \sqrt{k^+_1 k^+_2}} 
f(k^+_1) f(k^+_2) $\\

\blft$ \{ g^{++}_2 (k_1,k_2 ; x) [\,a^+(k^+_1) a^+(k^+_2) e^{-{i \over 2}(k_1+k_2).x}$\\

\blft$+a(k^+_1)a(k^+_2) e^{{i \over 2}(k_1+k_2).x}\,] $
\beq
\label{eqphi2}
 + G_2 (k_1, -k_2 ; x)  a^+(k^+_1) a(k^+_2) e^{{i \over2}(k_2-k_1).x} \}. 
\eeq
The x-dependence in the amplitudes $g^{++}_{2}$ and $G_{2}=g^{+-}_{2}+g^{-+}_{2}$ is
induced by a non-translational invariant classical background field $\Phi _{c}\left( x\right)$.
It comes in addition to the free-field dependencies - present in the exponential
functions - as a non-perturbative phenomenon.

\mysection{Dynamical Zero Mode (DZM) in  $\psi _{2}$ }

We consider the case of a periodic classical background field in the long wavelength
limit (which we need for the linearization of the equations of motion) :
\beq
\label{bckf}
\Phi_{c_{periodic}}\left( x\right) = \widetilde{\Phi}_{c}\left( k\right) e^{-\frac{i}{2}k.x} 
\eeq
Due to the linearization assumption in $\widetilde{\Phi}\left(k\right)$ the background field imposes 
a periodic variation in $x$ on $g^{++}_{2}$ and $G_{2}$ via the equations of motion. Therefore the following form is assumed for the amplitude of the DZM mode:
\beq
\label{DZMG}
G_{2_{periodic}}\left( k^{+}_{1},-k^{+}_{2}; x\right)=G^{0}_{2}\left( k^{+}_{1},-k^{+}_{2}\right) e^{-\frac{i}{2}k x}
\eeq
The resulting form of zero mode, which is noted $\hat{\Omega}_{k}$, is obtained from Eq.(\ref{eqphi2}) by integration of $\psi_{2}$ over $x^{-}$. It takes the general form
\beq
\label{omegak}
\hat{\Omega}_{k}=\int dk^{+}_{1} C_{1}(k^{+}_{1},k^{+})a^{+}(k^{+}_{1}+k^{+})a(k^{+}_{1}),
\eeq 
where the coefficient $C_{1}(k^{+}_{1},k^{+})$ is related to $G^{0}_{2}(k^{+}_{1},-k^{+}_{1}-k^{+}_{2})$.
For $k^{+} = 0$ the usual zero mode component of $\psi_{2}$ is obtained \cite{GUW}. $\hat{\Omega}_{k}$ is diagonal in the number of particles but it can transfer momentum $\pm k^{+}$, though
it is translationally invariant. This property is a consequence of motion in
a periodic background field.
\mysection{Solution of the Coupled Equations for $\Phi_{c}$, $g^{++}_{2}$ and $G_{2}$.}
Only linear terms in  $\Phi_c(x)$ are kept in the EQM which, after projection onto the vacuum, one and two-particles states, gives a coupled system for $\Phi_c(x), G_2$ and $g^{++}_2$. We only give 
the general structure for the Fourier components, $\tilde{\Phi}_c, \tilde{g}^{++}_2$ and $ \tilde{G}_2$,
with a renormalized squared-mass $\mu^2=m^2 + {\lambda \over 8\pi} \int^{\infty}_0  {dk^+ \over k^+} f^2(k^+)$
\beq
\label{A}
\Delta ^{-1}_{\Phi }\, \widetilde{\Phi _{c}}+K_{\Phi }\otimes \left[\tilde{G}_{2}\oplus \,\tilde{g}^{++}_{2}\right] =0, 
\eeq
\beq
\label{B}
\Delta ^{-1}_{G_{2}}\, \tilde{G}_{2} + \, K^{+-}\, \otimes \left[\tilde{G}_{2}\oplus \,\tilde{g}^{++}_{2}\right] =g\, \widetilde{\Phi _{c}}, 
\eeq
\beq
\label{C}
\Delta ^{-1}_{g_{2}}\, \tilde{g}^{++}_{2} + \, K^{++}\, \otimes \left[\tilde{G}_{2}\oplus \,\tilde{g}^{++}_{2}\right] =g\, \widetilde{\Phi _{c}}. 
\eeq
Here the $\Delta^{-1}$'s are inverse propagators for the fields on which they act and the K's are interaction kernels. $\tilde{\Phi}_c(k^+, k^-)$ is the driving term for $\tilde{G}_2$ and $\tilde{g}^{++}_2$  which are
therefore truely nonperturbative quantities.  However one has to distinguish between the two cases $k^+ = 0$ and $k^+ \ne 0$.
The first case yields the equation for critical coupling and the second the covariant
dispersion relation.

\subsection{Critical coupling $(k^+ = 0)$}
\vspace*{0.4cm}
The starting point is the analysis of Ref \cite{GUW}. With $C(q) \equiv C_{1}(q,0)$ in Eq.(\ref{omegak}) the equations determining the critical coupling read\\

\blft$C(q_1)[\,q_1 + g \,] + g \phi_0 + 2 g \int_{0}^{\infty} {dk_1 \over k_1} f^2(k_1)\{\tilde{g}^{++}_2 (k_1, q_1) $\\
\beq
\label{eqC0q}
 + {1\over 4}\,[\,\tilde{G}_2(k_1,- q_1)  + \tilde{G}_2(q_1,- k_1)\,]\}=0,  
\eeq
\par
\par
\blft$\phi_0+{g \over 6} \int{dk_1 \over k_1}f^2(k_1)\{f^2(k_1) C(k_1) + \int {dk_2 \over k_2} $
\beq
\label{eqgcr}
f^2(k_2)[\,4\, \tilde{g}^{++}_2( k_1,k_2)+ \tilde{G}_2(k_1,-k_2)\,]\} = 0,
\eeq
where $g={\lambda \over 4 \pi \mu^2}$ is the dimensionless coupling. The order parameter is $\phi_0 \equiv \tilde{\Phi}_{c}(0)$, the constant value of the classical field. The simplest approximate solutions to these equations  neglects the terms in $\tilde{g}^{++}_2$ and  $\tilde{G}_2$. Eq.(\ref{eqC0q}) gives then
\beq
\label{C0q}
C^{(0)}(g,q) = - {{g \phi_{0}} \over {q_1+g}}.
\eeq
For $\phi_{0} \neq 0$ Eq.(\ref{eqgcr}) determines the critical coupling $g^{(0)}_{cr}$ after proper renormalization
\cite{GUW}. With (\ref{C0q}) Eq.(\ref{eqgcr}) becomes
\beq
\label{eqgcr0}
\phi_{0} [1-{g^{(0)}_{cr} \over 6 } log(g^{(0)}_{cr})]=0 \,\,\,\,i.e.\,\,\,g^{(0)}_{cr}=4.19. 
\eeq
To extend this simple analysis an iterative procedure is used first to study the properties of the solutions to the system of Eqs.(\ref{B}-\ref{C}). The zero mode contributions
 in the kernels $K$ of these equations are isolated. This fixes the initial source functions
 $\tilde{g}^{(0)}_{2}$ and $\tilde{G}^{(0)}_{2}$  in the iterative procedure. It is found that the iterated solutions keep the same shape as the initial ones and that, oder by order, the changes occur only in a redefinition of the coupling 
$g$ present in $C^{(0)}(g,q)$ of Eq.(\ref{C0q}). This final effective coupling $g_f$ results from the resummation of a geometric series in $-g x(g) {{\sqrt{3} \pi} \over 9}$
\beq
\label{gf}
g_f={g \over {1+g x(g) {{\sqrt{3} \pi} \over 9}}},
\eeq
with $x(g)=1+{g \over 12}$ for $0 \leq g \leq 7$. The approximate solution $\tilde{g}^{++}_2$ 
 finally reads\\

\blft$\tilde{g}^{++}_2(q_1,q_2)={g \over 4}({{q_1 q_2} \over {q_1^2+q_2^2+q_1 q_2}})[C^0(g_f,q_1) f^2(q_1)$
\beq
\label{g2++}
+C^0(g_f,q_2) f^2(q_2) +2 \phi_0],
\eeq
while one has simply $\tilde{G}_{2}(q_1,-q_2)=\theta(q_1-q_2){\tilde{g}}^{++}_2(q_1,-q_2)$. Going back to Eq.(\ref{eqgcr})
and after renormalisation the equation for $g_{cr}$ is found as\\

\blft$1 - {g_{cr} \over 6} ( 1 + {2 \over 9} g_{cr} \pi \sqrt{3}) \ln [g_{cr}]$
\beq
\label{gcreq}
 + {1 \over 27} g_{cr}^2 \pi \sqrt{3} \ln[1 + {1 \over 9} g_{cr} \pi \sqrt{3} (1 + {g_{cr} \over 12})] = 0, 
\eeq
which gives $ g_{cr}=4.78$. For comparison with other studies using the convention of Parisi et al \cite{Parisi}
 this has to be translated \cite{GUW} into the reduced coupling unit $r$ which is just 1 at the perturbative
one loop level. To the above values of $g^{(0)}_{cr}$ and $g_{cr}$ corresponds the values $r = 1.5$ and
$r = 1.71$. This is in agreement with recent estimates from  high-temperature ($r=1.754$) \cite{BuCo,PeVi} and
 strong coupling expansions ($r=1.746$) \cite{CaPe} and  Monte Carlo simulations ($r=1.71$) \cite{KiPa,Kim}.
A recent high precision estimate \cite{CaHa} of $g_4$ for the 2D Ising model leads to $r = {3 g_4 \over 8 \pi} = 1.7543$. However the RG-improved fifth order perturbative result of Ref. \cite{OrSo} is $r_5 = 1.837(30) $. For a discussion of this
result see Ref.\cite{CCCPV}.

\subsection{Dispersion relation $(k^+ \neq 0)$}
\vspace*{0.4cm}
The DZM of Eq.(\ref{omegak})is now explicitely isolated in Eq.(\ref{B}). A generalization of Eq.(\ref{eqC0q})
for the coefficient $C_1(q_1^+,k^+)$ follows: upon neglecting the integral contributions in $\tilde{g}^{++}_2$ and  $\tilde{G}_2$, it reads simply\\

\blft$\{ 1 + {g \over 2 V}[{1 \over q_1^+}+{1 \over q_1^++k^+}]\} C_1(q_1^+,k^+)$
\beq
\label{eqC1q}
+{{g \phi_c(k)} \over {V\sqrt{q_1^+(q_1^++k^+)}}} =0,
\eeq
where $V$ is the volume of the invariant measure $[dx^{-}]$. Its specific analysis is discussed below. The solution is then
\beq
\label{C1q}
 C_1(q_1^+,k^+)\!=\!{{-g \phi_c(k)\sqrt{q_1^+(q_1^++k^+)}} \over {Vq_1^+(q_1^++k^+)+{g \over 2}(2q_1^++k^+)}}
\eeq
With this expression for $C_1$ used in  Eq.(\ref{A}) the dispersion relation becomes

$\blft[\mu^2-k^2]\phi_c(k) + {\lambda \over 24 \pi} \int_{0}^{\infty} dk_1^+[{f^2(k_1^+)f^2(k_1^++k^+) \over 
\sqrt{k_1^+(k_1^++k^+)}}$
\beq
\label{disper}
 C_1(k_1^+,k^+)] +\mbox {  non-zero modes terms} = 0.
\eeq
Given this form the question of covariance of the DZM contribution immediately arises. This issue  can be  
clarified with the help of the  $2nd$ order perturbative contribution to the self-energy (sunset diagram).
Indeed its explicit form written in terms of light-cone momenta is not evidently covariant but becomes so
if momenta are expressed in units of the external momentum $k^+$. Following the same path, in Eq.(\ref{C1q}), with $q_1^+$ expressed in units of $k^+$, the volume $V$, which has dimension of length, must also be written in units of $1 \over k^+$, {\it i.e.} V is just ${1 \over k^+}$ times a function of  Lorentz scalars. What is its argument? The external source function $\Phi_c(x)$ actually serves to probe characteristic distances of the system. For a periodic 
background field the interaction provokes a dispersion which induces another length scale, the energy
 flow scale, which is related to the off-shellness of the process. It is measured in terms of $k^-$, 
say $k^- \over \mu^2$. Hence in effect $V  \propto  ({1 \over k^+})^{\alpha}({k^- \over \mu^2})^{(1-\alpha)}={1 \over k^+}({k^2 \over \mu^2})^{(1-\alpha)}$. Since $\alpha$ is arbitrary any scalar function 
of ${k^2 \over \mu^2}$ is legitimate. Thus 
\beq
V={1 \over k^+} {\it v}({k^2 \over \mu^2}) \equiv {k^- \over \mu^2} {\it w}({\mu^2 \over k^2}),
\eeq
with ${\it v}(z) = z {\it w}(1/z)$. The infinite volume limit is then performed such that $(k^+ \to 
0, k^- \to \infty)$  with $k^2$ fixed. $C_1(q_1^+,k^+)$ of Eq.(\ref{C1q}) is therefore finite in the infinite 
volume limit and the DZM contribution to the dispersion relation can be written  $I_{DZM}(k^2)\phi_c(k)$ with $I_{DZM}(k^2)$ given as\\

\blft$I_{DZM}(k^2)=-{\lambda g \over 24 \pi} \int_0^\infty dx {1 \over {\it v}({k^2 \over\mu^2})x(1+x) + {g \over 2}(2x+1)}$
\beq
\label{dispint}
=-{\lambda g \over 24 \pi} {1 \over h(g,v)} 
\ln \Big[{{g+{\it v}({k^2 \over\mu^2})+h(g,v)} \over {g+{\it v}({k^2 \over\mu^2})-h(g,v)}}\Big],
\eeq
with $h(g,v)=\sqrt{g^2+{\it v}^{2}({k^2 \over\mu^2})}$. The dispersion relation becomes simply
\beq
k^2-\mu^2-I_{DZM}(k^2) = 0.
\eeq
In the limit $k^2 \to 0$ it should reproduce the constraint (after renormalization)
\beq
\label{theta3}
\theta_3= \phi_c \mu^2(1-{g \over 6} \ln (g)) = 0.
\eeq
The comparaison of these last two relations shows that $\lim_{k^2 \to 0} {\it v}({k^2 \over \mu^2})$
should be zero. For ${\it v}({k^2 \over \mu^2})\!\ll\!g$ the following expansion holds\\ 

\blft${1 \over \mu^2} I_{DZM}(k^2)\!=\!-{g \over 6} [\,\ln (g)\!-\!\ln ({{\it v} \over 2})\!+\!{{\it v}
\over g}]\!+\!{\it O}({\it v}^2 \ln ({\it v})).$\\

In this expression the divergent part is given by the $\ln ({\it v})$ term and is taken care of by renormalisation.
If $\lim_{k^2 \to 0} {\it v}({k^2 \over \mu^2}) \propto (k^2)^\alpha$ with $\alpha < 1$  the leading term in the
dispersion relation is the one linear in ${\it v}$. Hence the dependence of ${\it v}$ on $k^2$ determines completely the dispersion relation\footnote{At D=2 all perturbative contributions to $\Gamma (k^2)$ are  analytic in $k^2$.} and therefore the critical exponent $\eta$. The scaling behaviour would just be a pure volume effect. However the precise expression of $I_{DZM}(k^2)$ depends on the power of the scalar field interaction term $\Phi^{2 k}$, through the form of the constraint determining $C_1(q_1^+,k^+)$. The  form of ${\it v}({k^2 \over\mu^2})$  is dictated \cite{SGW} by conformal transformations preserving the local light cone structure. It is well known that  critical exponants of the $2D$ Ising model just come out from conformal invariance of the underlying fermionic field theory \cite{YG}.\\

It is quite instructive to make a comparison with the would be situation in DLCQ. Instead of the integral 
in Eq.(\ref{dispint}) the following sum is obtained
\beq
S(m,g) = \sum \limits_{n=1}^{\infty}{1 \over n(n+m)+{g \over 2}(2n+m)},
\eeq
where m is the mode number of the external $k^+={2 \pi m \over L}$. The sum is finite and yields different 
discrete results for each possible value of $m = 0, 1, 2,...$. It is boost invariant and independant of the size $L$. However the limit ${k^2 \to 0}$, which is necessary for the extraction of $\eta$, cannot 
be performed. One might think to repair this by an analytic prolongation to the domain $0<m<1$. But this has to be done keeping $k^2$ fixed. Moreover, if $k^{+}(m)$ is not fixed anymore by periodic boundary conditions, 
boost invariance is lost and has to be restored in one way or another. But this goes beyond the DLCQ framework, raising questions of consistency. On the contrary CLCQ permits a consistent analysis in the limit ${k^2 \to 0}$.\\

\mysection{$\beta$-function and critical exponent $\omega$ }

To determine the $\beta$-function the constraint is considered as a prescription for
the calculation of the critical mass ({\it cf} Eq.(\ref{theta3})). $M^2(g, \Lambda)$
is just $\mu^2$ times the left hand side of Eq.(\ref{gcreq}), with $\mu^2 = \lambda/(4 \pi g)$. One has
\beq
\label{betag}
\beta(g) = M [{\partial M \over \partial g}]^{-1}_{(\lambda, \Lambda)}=-2g{N(g) \over D(g)}.
\eeq
with\\ 

\blft$N(g)\!\!=\!\![{1-{g \over 6}(1+g{2\pi \sqrt{3} \over 9})\ln(g)}+$\\ 

\blft$g^2 {\pi \sqrt{3} \over 27}\ln(1 + g {\pi \sqrt{3} \over 9} (1 + {g \over 12}))][1 + g {\pi \sqrt{3} \over 9} (1 + {g \over 12})]$\\
 
and\\

\blft$D(g)=(1+{g \over 6}) [1+g{ \pi \sqrt{3} \over 9} (1+{g \over 12})]+$\\ 

\blft${g^2 \pi \sqrt{3} \over 27} (1 - { g^2 \pi \sqrt{3} \over 108}) 
+{g^2 \pi \sqrt{3} \over 27} [1 + {g\pi \sqrt{3} \over 9} (1 + {g \over 12})]$\\

\blft$\ln ({g \over (1 + g {\pi \sqrt{3} \over 9} (1 + {g\over 12}))})$\\
 
From standard renormalization group analysis one must have\\
\beq
\lim_{g \to 0} \beta(g) =-(4 - D)g+O(g^2)
\eeq
which is indeed the case for (\ref{betag}). To lowest order in the Haag expansion the
corresponding $\beta$-function is just\\
\beq
\label{beta0g}
\beta_0(g)=-{2g(1 - {g \over 6} \ln g) \over (1 + {g \over 6})}.
\eeq
The two functions $\beta(g)$ and $\beta_0(g)$  are plotted in Fig.1. In both cases the exponent $\omega=\beta^{\,\prime}(g_{cr})=2$ exactly, in agreement 
with the recent analysis of Ref.(\cite{CCCPV}). In particular these authors emphasize that at $D=2$ perturbative estimates of the critical value of $g$ and  of $\omega$  are not reliable due to strong nonanalytic corrections to the $\beta$-function.\\
\newpage
\begin{figure}[htb]
\hspace*{0.25cm}\psfig{file=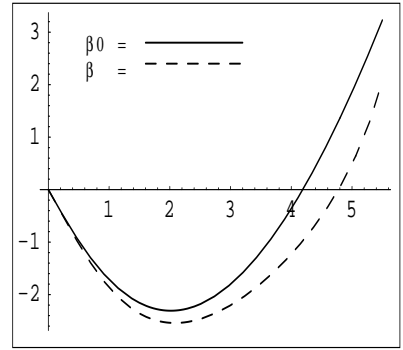,height=8.5cm,width=6.0cm}
\vspace*{-1.5cm}\caption{The functions $\beta(g)$, Eqs.(\ref{betag}-\ref{beta0g}).}
\label{fig:betag}
\end{figure}

\vspace{0cm}

\mysection{Conclusion}

Critical behaviour in LC quantization is a key issue since it is the domain of non-pertubative physics, an alleged benefit of this particular quantization scheme. However the
problem of scaling behaviour ({\it{e.g.} critical exponents)} has so far attracted little interest as compared to mechanisms of symmetry breaking. Whereas discretized LC techniques can still be useful for symmetry breaking studies, we have shown here that a full understanding of the physics of zero modes combined with continuous non-compact field dynamics is mandatory to treat covariantly the low $k^+$ region determining the scaling behaviour. Conformal transformations preserving the local light cone structure are
then found to govern precisely the scaling law. Yet it is a puzzling problem
to relate directly the CLCQ approach to $2D$ conformal field theory where critical exponents are known to be exactly determined by conformal symmetry properties. Finally
the $\beta$ function, although in keeping with RG analysis at small coupling,
presents interesting non-analytic properties and leads to a critical exponent $\omega=2$
in agreement with recent non-perturbative analysis.

\end{document}